\documentclass{article}
\usepackage{ijcai19}

\usepackage{times}
\usepackage{soul}
\usepackage{url}
\usepackage[hidelinks]{hyperref}
\usepackage[utf8]{inputenc}
\usepackage[small]{caption}
\usepackage{graphicx}
\usepackage{amsmath}
\usepackage{amsfonts,amssymb}
\usepackage{booktabs}
\usepackage{subfig}

\title{Quaternion Collaborative Filtering for Recommendation}

\author{
Shuai Zhang$^1$\footnote{Contact Author}\and
Lina Yao$^1$\and
Lucas Vinh Tran$^2$\and
Aston Zhang$^3$\And
Yi Tay$^2$\\
\affiliations
$^1$University of New South Wales\\
$^2$Nanyang Technological University\\
$^3$Amazon Inc.\\
\emails
\{shuai.zhang@student., lina.yao@\}@unsw.edu.au, \\
\{trandang001, ytay017\}@e.ntu.edu.sg,
astonz@amazon.com
}

\begin{document}

\maketitle

\begin{abstract}
This paper proposes Quaternion Collaborative Filtering (QCF), a novel representation learning method for recommendation. Our proposed QCF relies on and exploits computation with Quaternion algebra, benefiting from the expressiveness and rich representation learning capability of Hamilton products. Quaternion representations, based on hypercomplex numbers, enable rich inter-latent dependencies between imaginary components. This encourages intricate relations to be captured when learning user-item interactions, serving as a strong inductive bias  as compared with the real-space inner product. All in all, we conduct extensive experiments on six real-world datasets, demonstrating the effectiveness of Quaternion algebra in recommender systems. The results exhibit that QCF outperforms a wide spectrum of strong neural baselines on all datasets. Ablative experiments confirm the effectiveness of Hamilton-based composition over multi-embedding composition in real space.

\end{abstract}

\section{Introduction}
With the significant rise in the amount of information and number of users on the internet, it is becoming important for companies to provide personalized recommendations. As such, recommendation engine becomes an indispensable component to modern e-commerce. To this end, an effective recommendation model can not only significantly boost revenues, but also improve the overall user experience by ameliorating the prevalent and intrinsic problem of over-choice.

Learning a matching function between user and item lives at the heart of modern methods for recommender systems research. More concretely, factorization-based methods \cite{koren2008factorization,he2017neural}, metric learning methods \cite{hsieh2017collaborative,tay2018latent,zhang2018next}  and/or neural network models \cite{zhang2017deep,he2017neural} have all recently demonstrated good progress and performance on the task at hand. All in all, joint representation learning techniques forms the crux of this paradigm and investigating novel methods for this purpose remains highly interesting and relevant.

A key observation is that most work in this area has been primarily focused on real-valued representations $\mathbb{R}$, ignoring the rich potential of alternative spaces such as complex $\mathbb{C}$ and/or hypercomplex spaces $\mathbb{H}$.
This work investigates the notion of complex algebra and quaternion algebra which are well-developed in the area of mathematics. The intuition is clear. Complex and hypercomplex representation learning methods are not only concerned with expanding the vector space (i.e., merely composing multiple spaces together) but have tight links with associative retrieval, asymmetry and learning latent inter-dependencies between components via multiplication of complex numbers and/or Hamilton products. The associative nature of complex representations going beyond multi-view representations is well-established in the literature \cite{hayashi2017equivalence,danihelka2016associative}. Moreover, the asymmetry of simple inner products in hypercomplex space \cite{trouillon2016complex,tay2018hermitian} provides a strong inductive bias for reflecting the asymmetrical problem of user-item matching, i.e., user and item embeddings fundamentally belong to a separate class of entities.

To this end, Quaternion representations are hypercomplex numbers with three imaginary numbers. There have been recent surge in interest,  showing promise in real world applications such as signal processing~\cite{witten2006quaternion}, image processing~\cite{DBLP:journals/corr/abs-1811-02656}, and speech recognition~\cite{Parcollet2018SpeechRW,DBLP:journals/corr/TrabelsiBSSSMRB17}. This is in similar spirit to multi-view representations, although the latent components are connected via the complex-number system. Moreover, the Hamilton product encodes interactions between imaginary and real components, enabling an expressive blend of components that forms the final representation. Given the interaction function lies fundamental to recommender system research, it is intuitive that Hamilton products are suitable choices for user-item representation learning.

\paragraph{Our Contributions}
Overall, the prime contributions of this paper can be summarized as follows:
\begin{itemize}
    \item We propose recommender systems in non-real spaces, leveraging rich and expressive complex number multiplication and Hamilton products to compose user-item pairs. Our proposed approaches, Complex collaborative filtering and Quaternion collaborative filtering (QCF), opens up a new distinct direction for collaborative-based/neural recommendation in non-real spaces.
    \item We conduct extensive experiments on a wide range of large diverse real-world datasets. Our results demonstrate that QCF achieves state-of-the-art performance.
    \item We thoroughly investigate and control for parameters, demonstrating that the power of QCF lies not in its expanded representation capability but its expressiveness. When controlling for parameters, QCF outperforms MF significantly. Ablative experiments that control for latent dimensions prove the efficacy of our approach.
\end{itemize}

\section{Related Work}
In this section, we identify and review former studies that are relevant to our work.

Firstly, our work is concerned with collaborative filtering, in which matrix factorization~\cite{koren2009matrix} remains a competitive baseline. A wide spectrum of variants have been proposed based on it. For example, timeSVD++~\cite{koren2008factorization} takes temporal dynamics and implicit feedback into account and achieves good performance on rating prediction. BPR~\cite{rendle2009bpr} is a pairwise learning personalized ranking method with matrix factorization. \cite{he2016fast} proposed an efficient matrix factorization based recommendation model with a non-uniform weight strategy, just to name a few. Recent years have witnessed a shift from traditional collaborative filtering to neural networks approaches, with exponentially increase in the amount of publications on this topic~\cite{zhang2017deep,tay2019holographic,zhang2019deeprec}. The achievements are inspiring and enlightening.  Many of the well-established neural networks can be applied to recommender systems. For instance, We can add nonlinear transformations into collaborative filtering with multilayer perceptron~\cite{dziugaite2015neural,he2017neural,zhang2017autosvd++,zhang2018neurec,tay2018multi}, use autoencoder or convolutional neural networks to learn richer feature representations from user/item auxiliary information (e.g., profiles, images, video abstract, reviews, etc.)~\cite{sedhain2015autorec,van2013deep,chen2017personalized}, and utilize recurrent neural networks to model the sequential patterns in user behaviours to make session-aware recommendation~\cite{hidasi2015session,wu2017recurrent}.

Secondly, our work is inspired by the widespread success of complex and Quaternion number across a myriad of domains.  \cite{DBLP:journals/corr/TrabelsiBSSSMRB17} devised some atomic components for complex-valued deep neural networks and built a complex-valued convolutional neural network for several computer vision and speech processing tasks. \cite{gaudet2018deep,parcollet2018quaternion2} proposed the building blocks for Quaternion convolution and tested it on image classification tasks.  \cite{parcollet2018quaternion} generalized Quaternion into recurrent neural networks and showed better performances than RNN/LSTM on a realistic application of automatic speech recognition. In addition, Quaternion MLP~\cite{parcollet2016quaternionnlp,zhang2019quaternion} also achieved promising results on spoken language understanding and knowledge graph embedding. Former studies suggest that complex and Quaternion have a richer representational capacity, and enjoy higher flexibility/generalization than real-valued based methods.

\section{Preliminaries}
In this section, we give a brief introduction to complex and Quaternion algebra which lie at the centre of the proposed method.

\subsection{Complex Algebra}
A complex number is a number that can be written as
$ a + b \textbf{i}$, where $\textbf{i}$ is the imaginary unit, and $a$ and $b$ are the real numbers. The backbone of the complex algebra is the imaginary unit which is the solution of $x^2 = -1$.

The multiplication of complex number follows the general rule that each part of the first complex number gets multiplied by each part of the second complex number. Thus we have:
\begin{equation}
    (a  + b \textbf{i}) (c + d \textbf{i}) = (ac - bd) + (bc + ad) \textbf{i}
\end{equation}
The multiplication of complex number occurs in a two-dimensional space and obeys the commutative law and the result is still a complex number.

\subsection{Quaternion Algebra}
Quaternion is an extension of complex number that operates on four-dimensional space. It consists of one real part and three imaginary parts. In this work, we mainly focus on the Hamilton's Quaternion which is defined as:
\begin{equation}
    \mathcal{H} := \{a + b\textbf{i} + c\textbf{j} + d\textbf{k}\  | a, b, c, d \in \mathbb{R} \}
\end{equation}
where $a$ is the real component and $\textbf{i}$, $\textbf{j}$, and $\textbf{k}$ satisfy the relations:
\begin{equation}
    \textbf{i}^2 = \textbf{j} ^ 2 = \textbf{k} ^ 2 = \textbf{ijk} = -1
\end{equation}
Such a definition can be
used to describe spatial rotations.

The Hamilton product of two Quaternions is determined by the products of the basis elements and the distributive law. Suppose that we have two Quaternions $\mathcal{H}_1={a_1 + b_1\textbf{i} + c_1 \textbf{j} + d_1\textbf{k}}$ and $\mathcal{H}_2={a_2 + b_2 \textbf{i} + c_2 \textbf{j} + d_2 \textbf{k}}$, the Hamilton product of them is defined as follows:
\begin{align}
\begin{split}
        \mathcal{H}_1 \otimes \mathcal{H}_2 &= (a_1 a_2 - b_1 b_2 -c_1 c_2 -d_1 d_2) \\&+ (a_1 b_2 + b_1 a_2 + c_1 d_2 - d_1 c_2 ) \textbf{i} \\&+ (a_1 c_2 -b_1 d_2 + c_1 a_2 + d_1 b_2 ) \textbf{j} \\&+ (a_1 d_2 + b_1 c_2 - c_1 b_2 +d_1 a_2 ) \textbf{k}
\end{split}
\end{align}
Multiplication of Quaternions is associative and distributes over vector addition, but it is not commutative.

\section{The Proposed Approach}
In this section, we propose applying complex and Quaternion number to recommendation models.

Suppose that we have $N$ items and $M$ users, user's preferences over items formulate a user-item \textit{interaction matrix} $Y \in \mathcal{R}^{M \times N}$. Our task is concerned with learning personalized ranking list with implicit interactions, e.g., click, like, check-in, watch, purchase, rate, etc. We set $Y_{ui}$ to $1$ if user $u$ has interacted with item $i$ and $Y_{ui}=0$ indicates that the user dislikes or does not know the existence of item $i$. Let $Y^+$ denote the user-item set with $Y_{ui}=1$ and $Y^-$ denote the set with $Y_{ui}=0$. The aim is to fill in the missing values in the interaction matrix.

\subsection{Collaborative Filtering with Complex Embedding}
In classic matrix factorization~\cite{koren2009matrix}, each user is represented by a latent vector of dimension $d$ and each item is also allocated with a vector with the same dimension. In the complex number system, we embed users and items with complex vectors. Formally, we define each user with a complex vector $\mathcal{C}_u = U_u + V_u \textbf{i}$ with $U_u, V_u \in \mathbb{R}^d$. Similarly, each item is represented with a complex vector $\mathcal{C}_i = P_i + Q_i \textbf{i}$ with $ P_i, Q_i \in \mathbb{R}^d$.

Different from traditional matrix factorization where dot product is usually used, we model the interaction between users and items with complex number multiplication, which gives the following expression:
\begin{equation}
    (U_u \cdot P_i - V_u \cdot Q_i) + (V_u \cdot P_i + U_u \cdot Q_i) \textbf{i}
\end{equation}
where dot product ``$\cdot$" is used in order to get a scalar output but it will not change the multiplication rule. For simplicity, we define by the real part of the above expression with $a$ and the imagery coefficient with $b$. Following~\cite{parcollet2018quaternion}, we apply split activation functions (specifically, \textit{sigmoid}) to the above result and get:
\begin{equation}
   \alpha(a) + \alpha(b) \textbf{i} = \frac{1}{1 + e^{-a}}+ \frac{1}{1 + e^{-b}} \textbf{i}
\end{equation}
We use split activation because it can lead to better stability and simpler computation~\cite{parcollet2018quaternion,gaudet2018deep}.

\subsection{Collaborative Filtering with Quaternion Embedding}
Similar to complex number, we embed users and items with Quaternion representations. In formal, let $\mathcal{H}_u = U_u + V_u \textbf{i} + X_u \textbf{j} + Y_u \textbf{k}$ denote each user and $\mathcal{H}_i = P_i + Q_i \textbf{i} + S_i \textbf{j} + T_i \textbf{k}$ denote each item, where $U_u, V_u, X_u, Y_u, P_i, Q_i, S_i, T_i \in \mathbb{R}^d$. We model the user item relationship with the well-defined Hamilton product and obtain:
\begin{align}
\label{qcf}
\begin{split}
      &   (U_u \cdot P_i - V_u \cdot Q_i -X_u \cdot S_i -Y_u\cdot T_i) \\&
      + (U_u\cdot Q_i + V_u \cdot P_i + X_u \cdot T_i - Y_u \cdot S_i ) \textbf{i} \\&
      + (U_u\cdot S_i -V_u \cdot T_i + X_u \cdot P_i + Y_u \cdot Q_i ) \textbf{j}\\&
      + (U_u \cdot T_i + V_u \cdot S_i - X_u\cdot Q_i +Y_u \cdot P_i ) \textbf{k}
\end{split}
\end{align}
Note that dot product ``$\cdot$" is also used to get a scalar result. As can be seen, Hamilton product enables interactions between different components, thus, leading to more powerful modelling capability. Additionally, the noncommutative property also enables the recommender systems to model both symmetrical (when all imaginary parts are zero) and asymmetrical collaborative effects. Moreover, Hypercomplex spaces enable spatial transformations (and smooth rotations), as compared to real space products. This enables the model to be more numerically stable and expressive.

Similar to complex collaborative filtering, we use $a, b, c, d$ to represent the real and imaginary components of expression \eqref{qcf}. Split \textit{sigmoid} activation function is also employed over the result. Thus, we have:
\begin{equation}
    \alpha(a) + \alpha(b)\textbf{i} +\alpha(c)\textbf{j} + \alpha(d)\textbf{k}
\end{equation}

With complex or Quaternion embedding, we could not only model the internal interactions between user and item latent factors, but also external relations of different latent vectors. Intuitively, Quaternion Hamilton product is able to capture more intricate relations than complex number multiplication.

\subsection{Model Learning and Inference}
Following~\cite{parcollet2018quaternion,gaudet2018deep}, we train our model with classic cross-entropy loss in a component-wise manner. Here, we give the loss function for complex collaborative filtering. Loss function of Quaternion collaborative filtering can be easily derived as follows.
\begin{align}
\begin{split}
    \ell = &- \sum_{(u,i) \in \mathcal{Y}^+ \cup \mathcal{Y}^-} Y_{ui} \log(\alpha(a)) + (1 - Y_{ui}) \log(1 -\alpha(a)) \\ &- \sum_{(u,i) \in \mathcal{Y}^+ \cup \mathcal{Y}^-} Y_{ui} \log(\alpha(b)) + (1 - Y_{ui}) \log(1 -\alpha(b))
\end{split}
\end{align}
Where $ \mathcal{Y}^+ \in Y^+$ and $\mathcal{Y}^-$ is sampled from $Y^-$. Negative sampling is conducted during each training epoch with controlled sampling ratio. Same as NeuMF, we solve the personalized ranking task with implicit feedback from the classification perspective.  $\ell_2$ norm is used to regularize the model parameters. Note that pairwise Bayesian log loss~\cite{rendle2009bpr} is also viable for our model. Finally, the model can be optimized with standard gradient descent algorithms such as SGD or Adam optimizer~\cite{ruder2016overview}.

During the inference stage, we take the average of all the components as the final recommendation score. So the ranking score for complex collaborative filtering is:
\begin{equation}
    \hat{Y}_{ui} = \frac{\alpha(a) + \alpha(b)}{2}
\end{equation}
Similarly, the ranking score of Quaternion collaborative filtering is the mean of one real and three imaginary coefficients (with \textit{sigmoid} applied).
\begin{equation}
    \hat{Y}_{ui} = \frac{\alpha(a) + \alpha(b) + \alpha(c)+\alpha(d)}{4}
\end{equation}

With the mean aggregation, our approach ensembles predictions from different parts and, as a result, could reduce biases and variations, leading to a more robust estimation.

To accelerate the convergence and training efficiency,  we follow the  parameter initialization principle of~\cite{gaudet2018deep} which is tailored for Quaternions to initialize the proposed model.

\subsection{Quaternion Neural Network?}
Here, we give a brief discussion on Quaternion neural networks. Considering that our model follows the basic patterns (e.g., split activation function, component-wise loss, etc.) of Quaternion neural networks, it is easy to add some Quaternion feed-forward layers into our model~\cite{gaudet2018deep,parcollet2018quaternion2}. In Quaternion neural networks, the weights are also represented with Quaternions and transformation is performed with Hamilton product. The model learning and inference procedure remain unchanged. Complex neural networks share the same pattern as Quaternion ones. We omit details on this topic but will discuss it in Section 5.

\section{Experiments}
In this section, we present our experimental setup and evaluation results. Our experiments are designed to answer the following research questions:
\begin{enumerate}
    \item Does the proposed model outperform matrix factorization approach? How much is the improvement?
    \item Can our model beat recent neural network based model?
    \item How does the proposed model behave? For instance, where is the improvement from? What is the impact of hyper-parameters?
\end{enumerate}

\begin{table}[t]
\label{datasets}
\centering
\vspace{3mm}
\begin{tabular}{|l|c|c|c|c|}
\hline
Dataset & \#Users & \#Items & \#Action & Density\% \\ \hline

   ML HetRec     &   2.1k      &  10.1k      &  855.6k              &    4.005      \\ \hline
     LastFM &   1.0k     &  12.1k    &          83.4k      &         0.724 \\ \hline
      Foursquare &    7.4k     &  10.2k    &   101.4k             &      0.135   \\ \hline
      Video Games &  7.2k     &   16.3k  &    140.3k            &       0.119  \\ \hline
 Digital Music &   2.9k    &   13.2k  &    64.3k            &  0.169       \\ \hline
  Office Products &     2.4k  &        6.0k      &   44.4k    &      0.306   \\ \hline
\end{tabular}
\caption{Summary of the datasets in our experiments.}
\end{table}


\subsection{Dataset Description}
We conduct evaluation across a wide spectrum of datasets with diverse domains and densities. A concise summary is shown in Table 1.

 \begin{itemize}
     \item Movielens HetRec. It is a widely used benchmark dataset for recommendation model evaluations provided by GroupLens research\footnote{https://grouplens.org/datasets/movielens/}. We use the version HetRec.
     \item LastFM. This dataset is collected from the last.fm online music system\footnote{http://www.last.fm}. It contains music listening information of the website users.
     \item Foursquare. This is a location check-in dataset which records the digital footprints of users from the Foursquare application\footnote{https://foursquare.com/}.
     \item Amazon Datasets\footnote{http://jmcauley.ucsd.edu/data/amazon/}. This repository contains Amazon product ratings for a variety of categories. Due to limited space, we adopted three subcategories including: video games, digital music and office products.
 \end{itemize}
 For all datasets, we converted all interactions into binary values and ensured that each user has at least five interactions since cold-start problem is usually investigated separately~\cite{zhou2011functional}.

\subsection{Evaluation Setup and Metrics}
To evaluate the performance of the proposed approach, we adopted hit ratio (HR) and normalized discounted cumulative gain (NDCG) with top $k$ list. HR measures the accuracy of the recommendation model with definition $HR@k = \frac{\#\text{Hit}}{\#\text{Users}}$.  NDCG assesses the ranking quality of the recommendation list as, in practice, to make the items that interest target users rank higher will enhance the quality of recommendation lists. For each metric, we report the results with three different $k$ value, namely, top-5, top-10 and top-20. We report the average scores of both evaluation metrics. The higher both metrics are, the better recommendation performance is.

Following \cite{he2017neural}, we use the leave-one-out protocol to study the model behaviour. We hold out one interaction (the latest interaction if time-stamp is available) of each user as test and use the remaining data for training. During test step, we pair each test item with sampled $200$ (negative) items that the user have not interacted with. Then, we rank those $201$ items and report the evaluation results with the above metrics.

\subsection{Baselines}
To verify the performance of the proposed approach, we compare it with the following baselines.
\begin{itemize}
    \item Generalized Matrix Factorization(\textbf{GMF})~\cite{he2017neural}. It is a generalized version of matrix factorization and a special case of neural collaborative filtering framework. GMF learns its parameters by minimizing the cross-entropy loss.

    \item Multilayer Perceptron (\textbf{MLP}). It captures user and item relations with multilayered feedfoward networks. A three-layer MLP with pyramid structure is adopted.

    \item Joint Representation Learning (\textbf{JRL})~\cite{zhang2017joint}. JRL passes the element-wise product of user and item latent vectors into a tower-structured multilayered perceptron.
    \item Neural Collaborative Filtering (\textbf{NCF or NeuMF})~\cite{he2017neural}. NeuMF is an ensemble of MLP and GMF. MLP could help introduce non-linearity while GMF retains the advantages of matrix factorization. The whole framework is trained in an end-to-end manner with cross-entropy loss.

\end{itemize}
We denote by \textbf{CCF} the complex collaborative filtering and \textbf{QCF} the Quaternion collaborative filtering. Models such as ItemPOP and BPR are not further reported since they are outperformed by the above baselines~\cite{he2017neural}.

\subsection{Implementation Details}
All the above baselines and the proposed approaches are implemented with Tensorflow\footnote{https://www.tensorflow.org/}. We tested all models on a Titan Xp GPU. For fair comparison, pre-training is not used in the experiments. The regulariztion is tuned amongst $\{0.001, 0.005, 0.01, 0.05, 0.1\}$. The learning rate is selected from $\{0.001, 0.005, 0,01, 0.05\}$. The activation function $\alpha$ is set to \textit{sigmoid} function. We set latent vector dimension $d$ to the same value for all models for fair comparison. The batch size is fixed to $256$ in the experiment.

\begin{table*}[!ht]
\small
\centering
\label{results}
\begin{tabular}{|l|c|c|c|c|c|c|c|c|c|c|c|c|}
\hline

          & \multicolumn{6}{c|}{\textbf{Movielens HetRec}}                          & \multicolumn{6}{c|}{\textbf{LastFM}}                                          \\ \hline
          & \multicolumn{3}{c|}{Hit Ratio@k} & \multicolumn{3}{c|}{NDCG@k} & \multicolumn{3}{c|}{Hit Ratio@k} & \multicolumn{3}{c|}{NDCG@k} \\ \hline
          & k=5      & k=10      & k=20      & k=5     & k=10    & k=20    & k=5      & k=10      & k=20      & k=5     & k=10    & k=20    \\ \hline

MLP       &   0.2233       &       0.3473    &   0.5243        &   0.1424      &  0.1823       &    0.2267     &       0.4384   &    0.5783       &  0.6919         &  0.2690       &    0.3143     &    0.3432     \\ \hline
GMF       &  0.3024        & 0.4221          &    0.5575       &    0.1971     &   0.2355      &     0.2695    &    0.4921      &   0.5972        &  0.6982         &    0.3770     &   0.4107      &   0.4361      \\ \hline
JRL       &  0.2640      &  0.3814       &  0.5196        & 0.1812   &   0.2229     &     0.2568    &        0.4847 &     0.5814    &     0.6761     &   0.3573    &   0.3886     &      0.4126  \\ \hline
NCF     &    0.2621      &    0.4197       &   0.5844        &     0.1620    &       0.2125  &   0.2543      &  0.5368        &  0.6483         &    0.7261       &    0.3893     &    0.4256     &  0.4452       \\ \hline
\textbf{CCF} &    0.3066     &       0.4240    &       0.5650    &  0.1996      & 0.2376        &    0.2731     &     0.5015     &       0.5930    & 0.7160          &0.3933         &   0.4224      & 0.4539        \\ \hline
$\%^{*}$       &    +1.39   & +0.45      &   -3.32    &     +1.27   &  +0.89     &   +1.34  &  -6.57       &   -8.53       &   -1.39     &     +1.03  &    -0.75     &   +1.95     \\ \hline
\textbf{QCF}       &       0.3038   &       0.4477    &  0.5978         &  0.1957       &  0.2421      & 0.2798      &0.5450       &   0.6495        &  0.7413        &   0.4289     &   0.4626     &  0.4857    \\ \hline
$\%^{**}$       &    +0.46    &     +6.06  &  +2.29   &    -0.71    &  +2.81    &  +3.82     &      +1.53    &     +0.19      &    +2.09       & +10.17     &    +8.69    &   +9.09     \\ \hline

          & \multicolumn{6}{c|}{\textbf{Foursquare}}                          & \multicolumn{6}{c|}{\textbf{Video Game}}                                          \\ \hline
          & \multicolumn{3}{c|}{Hit Ratio@k} & \multicolumn{3}{c|}{NDCG@k} & \multicolumn{3}{c|}{Hit Ratio@k} & \multicolumn{3}{c|}{NDCG@k} \\ \hline
          & k=5      & k=10      & k=20      & k=5     & k=10    & k=20    & k=5      & k=10      & k=20      & k=5     & k=10    & k=20    \\ \hline

MLP       &   0.6316       &     0.7879      &    0.8443       &   0.3493      & 0.4015       &   0.4159      &  0.3554        &    0.4865       &   0.6254        &  0.2474       &     0.2899   &     0.3250    \\ \hline
GMF       &   0.7661       &  0.8204         &  0.8699         &   0.6453      &   0.6628      &   0.6754      &    0.2577      &  0.3831         &  0.4831         &   0.1727     &   0.2078     &  0.2378       \\ \hline
JRL       & 0.7369       &     0.7867    &  0.8270        &0.5377     &   0.5540     &  0.5642       &   0.2416      &    0.3488     &    0.4498      &  0.1578     &    0.1924    & 0.2180       \\ \hline
NCF     &        0.7657  &    0.8081       &   0.8369        &     0.5203    &     0.5343    &    0.5416     &   0.3593      &      0.4826     &  0.6042       &  0.2515      &     0.2912  &  0.3219      \\ \hline
\textbf{CCF} & 0.7891         &  0.8368        & 0.8765       &  0.6669      &     0.6823   &  0.6924      &     0.2958     &   0.4059        & 0.5293          & 0.2033        &   0.2388      &  0.2699       \\ \hline
$\%^{*}$       &      +3.00   &    +2.00    & +0.76      &   \textit{ +3.35}     &  +2.94   &   +2.52   &     -17.67     &   -16.57       & -15.37      &  -19.17       &    -17.99     &   -16.96      \\ \hline
\textbf{QCF}       &     0.7889     &   0.8312    & 0.8684          &    0.6523     &      0.6660   &  0.6754       &        0.4057  &  0.5323      &     0.6494     & 0.2869     &     0.3279 &   0.3577     \\ \hline
$\%^{**}$      &    +2.98     &    +1.32   &   -0.17   &   +0.70      & +0.48   &   +0.00  &       +12.91  &   +9.41      &      +3.84    &  +14.08       & +12.60       &  +10.06       \\ \hline

          & \multicolumn{6}{c|}{\textbf{Digital Music}}                          & \multicolumn{6}{c|}{\textbf{Office Products}}                                          \\ \hline
          & \multicolumn{3}{c|}{Hit Ratio@k} & \multicolumn{3}{c|}{NDCG@k} & \multicolumn{3}{c|}{Hit Ratio@k} & \multicolumn{3}{c|}{NDCG@k} \\ \hline
          & k=5      & k=10      & k=20      & k=5     & k=10    & k=20    & k=5      & k=10      & k=20      & k=5     & k=10    & k=20    \\ \hline
MLP       &    0.3301      & 0.4507          &  0.5665         &     0.2262    &     0.2655    &    0.2948     &      0.1324    &  0.2528         &  0.4470        &   0.0814      &     0.1194    &   0.1682      \\ \hline
GMF       &  0.2931        &    0.3944       &  0.5074         &  0.1973       &   0.2299      &    0.2585     &   0.1365       &    0.2735       &   0.5400        &     0.0798    &     0.1237    &    0.1906     \\ \hline
JRL       &   0.3349     &    0.4400     &   0.5578       &  0.2341   &    0.2680    &   0.2979      &    0.1706     &    0.3051     &   0.4902       &   0.1034    &   0.1462     &   0.1925     \\ \hline
NCF     &        0.3311  &      0.4303     &       0.5378    &     0.2322   &    0.2643     &   0.2915      &       0.1679  &     0.2723      & 0.4101        &  0.1066      &  0.1399     &   0.1746     \\ \hline
\textbf{CCF} &    0.3453      &   0.4497        &   0.5589        &    0.2361     &   0.2697      & 0.2974         &        0.1560  &      0.2955     &    0.5454       &   0.0972      &     0.1419    &     0.2046   \\ \hline
$\%^{*}$       &  +3.11     &     -0.22      &    -1.34      & +0.85        &  +0.63     &  -0.17    &  -8.56      &      -3.15    &   +1.00    & -8.84     &   -2.94   &   +6.29     \\ \hline
\textbf{QCF}       &   0.3821      & 0.4967         &0.6050          &  0.2662      &  0.3034      &   0.3309    &     0.1832    &  0.3038   &  0.4713      &0.1126  &    0.1511   & 0.1936      \\ \hline
$\%^{**}$       &    +14.09      &     +10.21     &      +6.79    &    +13.71    &   +13.21     & +11.08      &     +7.39     &   -0.43       &   -12.72        &      +5.64  &  +3.35     & +0.57       \\ \hline
\end{tabular}
\caption{Performance comparison between baselines and the proposed models: complex collaborative filtering (CCF) and Quaternion collaborative filtering (QCF).  We report hit ratio and NDCG at three cut off values. ``$\%^{*}$ " denotes the improvement of CCF over the strongest baseline; ``$\%^{**}$ " denotes the improvement of QCF over the strongest baseline.}
\end{table*}

\begin{figure*}[!ht]
\centering
\subfloat[]{\includegraphics[width=.6\columnwidth]{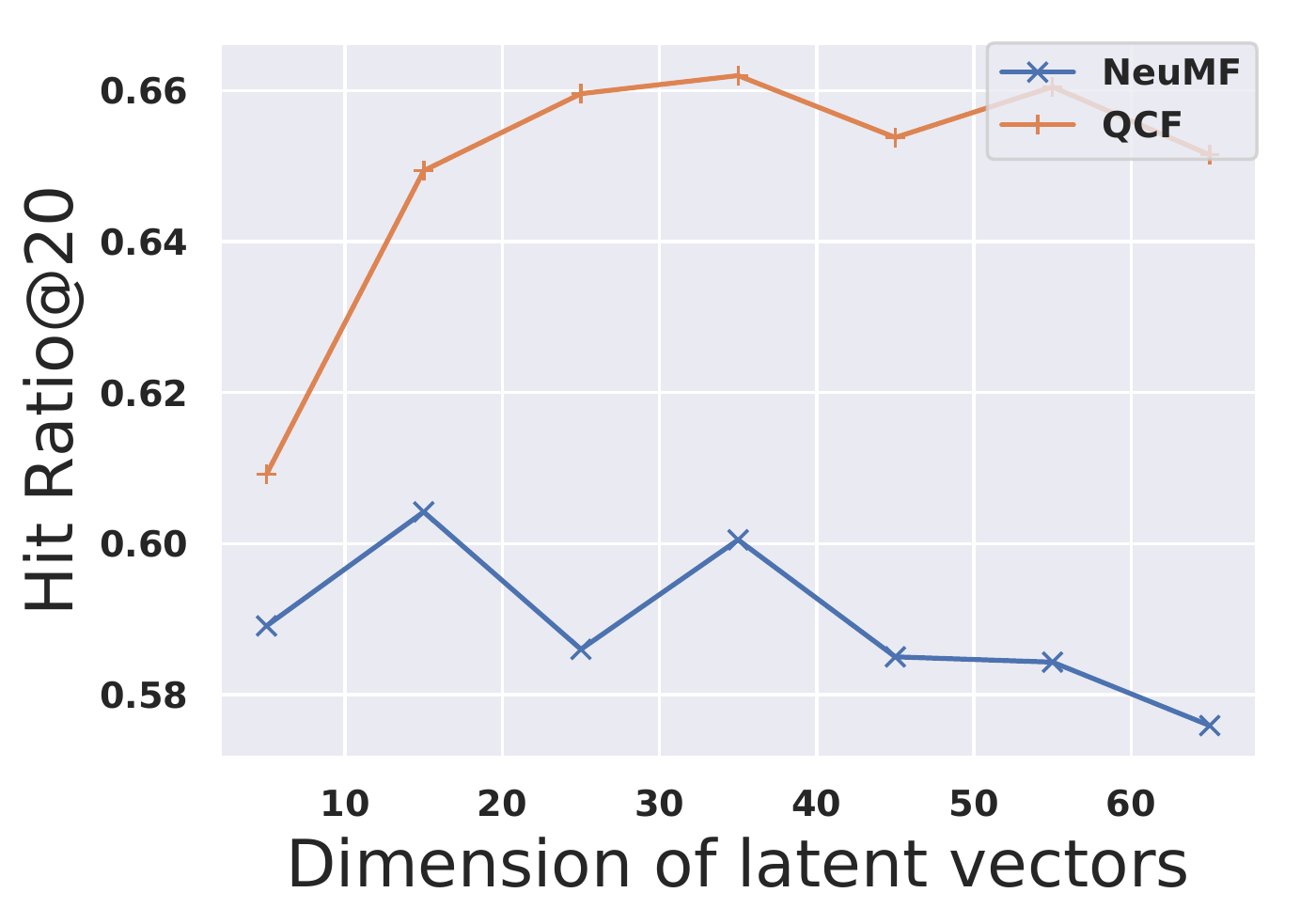}}
\quad
\subfloat[]{\includegraphics[width=.6\columnwidth]{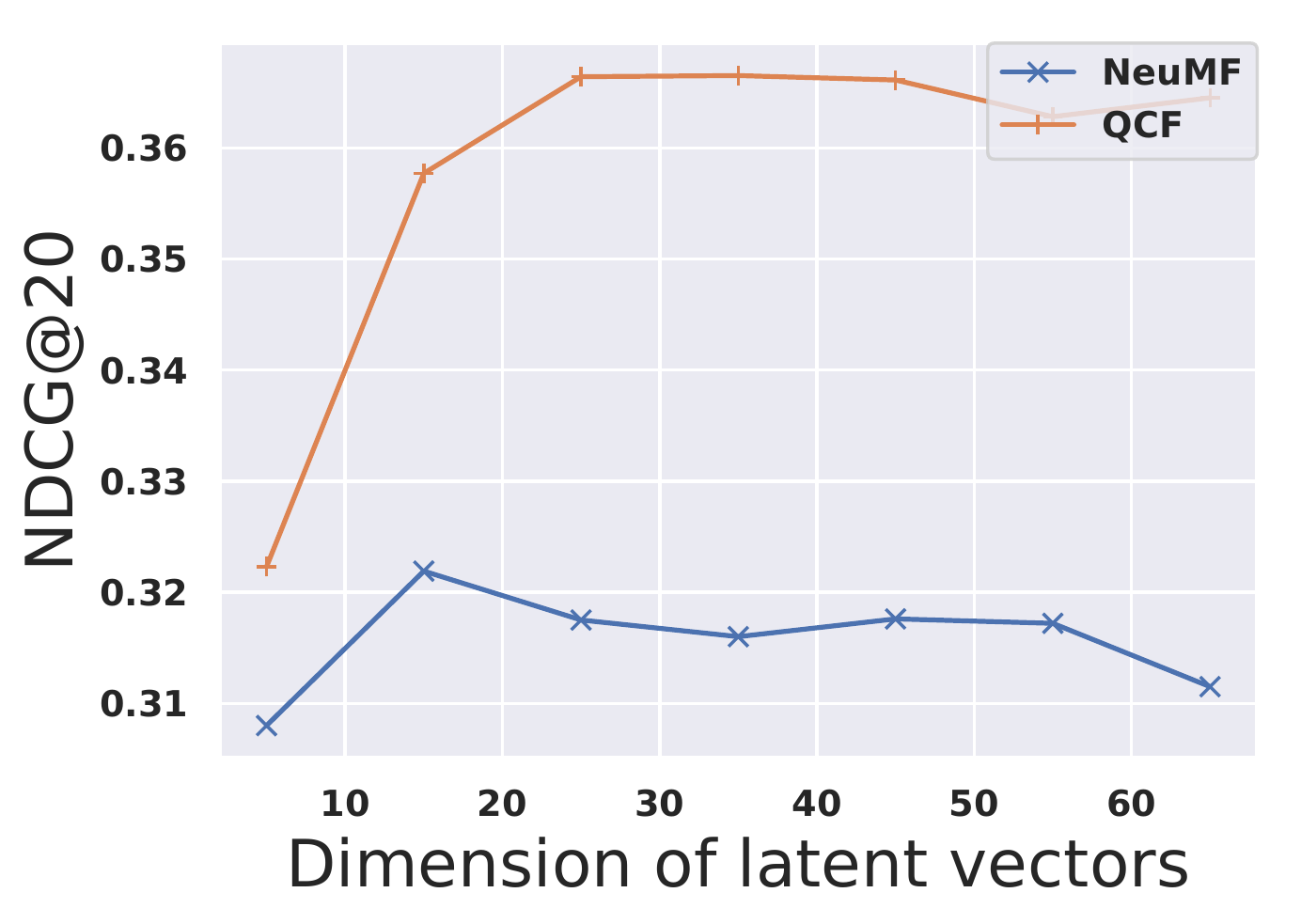}}
\quad
\subfloat[]{\includegraphics[width=.6\columnwidth]{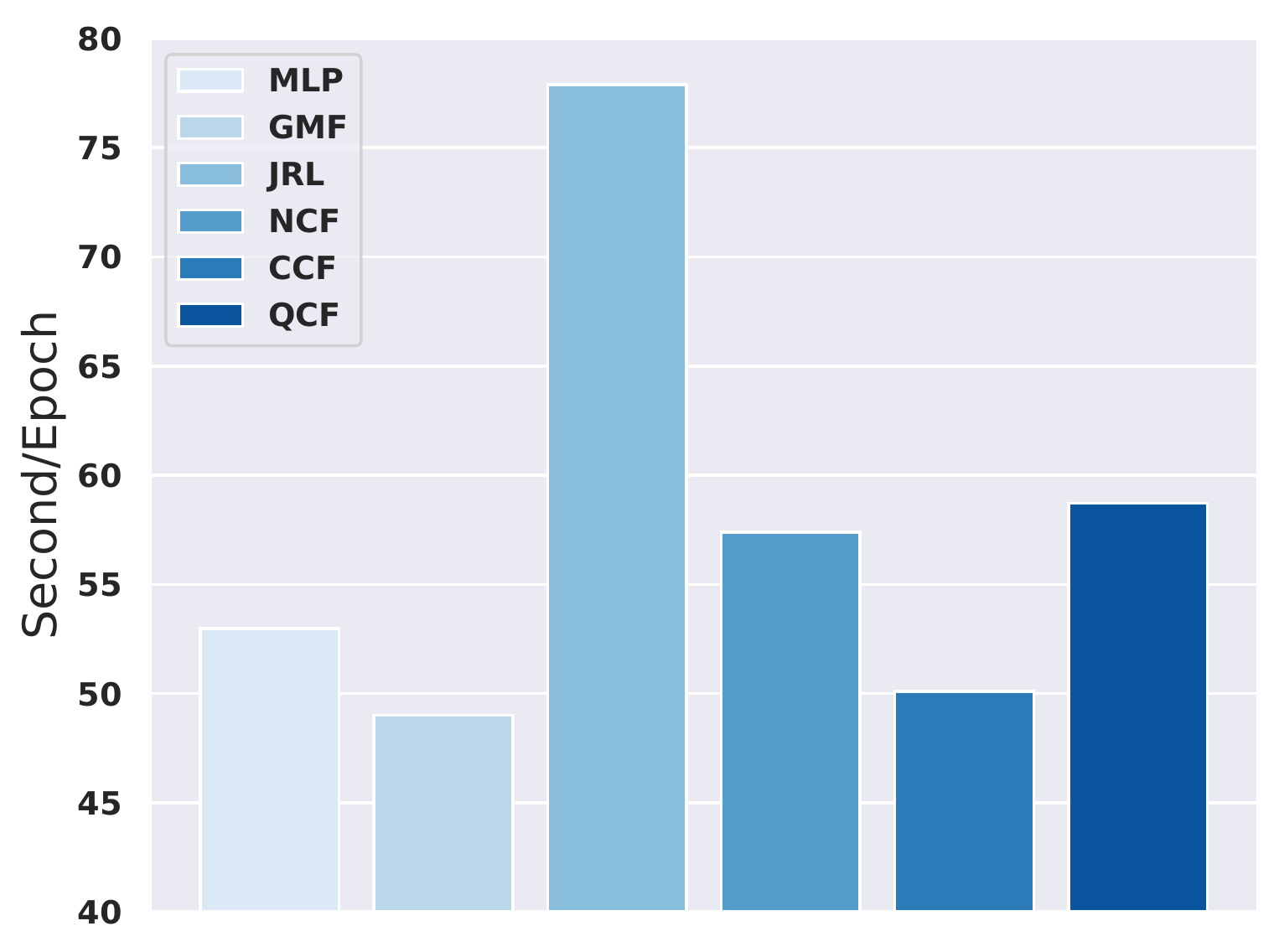}}
\caption{ Effects of latent vector dimension $d$ on datasets Video Game. We only report (a) HR@20 and (b) NDCG@20 due to limited space. (c) Runtime of six models on dataset Video Game. Experiment is run on NVIDIA TITAN X Pascal GPU.}
\label{fig:parame}
\end{figure*}

\subsection{Experimental Results}
The experimental results on six benchmark datasets are reported in Table 2. The relative improvement of CCF and QCF against the strongest baseline is also provided in the table.

Several observations can be made from the results. Firstly, the proposed QCF outperforms all other baselines.  The performance gain is large over the strongest baseline. On average, QCF improves the score of the strongest baseline by approximately $5.46\%$. It can not only get higher hit accuracy but also gain superior ranking quality. Secondly, we observe that there is no clear winner between CCF and the best baseline, but outperforms GMF in general. That is, CCF can achieve comparable performance to NeuMF even that it has simpler structures. Thirdly, CCF achieves slightly better performance than QCF on Foursquare, but fails on other five datasets with poorest performance on Video Game. In total, CCF is weaker compared with QCF. This is reasonable as Quaternion has better representational capability than complex number.  Fourthly, the hit ratio and NDCG usually show consistent patterns in terms of varying cut-off values.

In summary, the proposed approaches can outperform both matrix factorization and neural networks based recommendation models, which clearly answers research question one and two.

\subsection{Model Analysis}
To answer the research question three, we conduct model analysis to get a better understanding towards the proposed approaches. We mainly focus upon analyzing QCF since it usually performs the best.

\subsubsection{Where Does the Improvement Come From?} As we can see from Eq.\eqref{qcf}, each user/item has four latent embedding vectors. It is natural to assume that it is the increase of latent embedding vectors that leads to the performance improvement. In order to study the effect of the increase of embedding vectors, we devised an extension of GMF by using four embedding vectors for each user and item. For simplicity, we reuse the notations in Section 4.2 and use $U_u, V_u, X_u, Y_u \in \mathbb{R}^d $ to represent the user latent vectors and $P_i, Q_i, S_i, T_i  \in \mathbb{R}^d$ to denote item latent vectors. The prediction function for this model is formulated as:
\begin{equation}
    \sigma (U_u \cdot P_i + V_u \cdot Q_i + X_u \cdot S_i + Y_u \cdot T_i )
\end{equation}
We train the model following the settings (e.g., loss function, optimization algorithm, etc.) of generalized matrix factorization and name it ``MMF" (multiple matrix factorization).

The comparison results is shown in Table 3. At this point, we make an interesting observation - simply increasing the embedding vectors does not necessarily lead to substantial improvement and can even cause accuracy decrease. We can see that the improvement of MMF over GMF is trivial. We also found that it took longer time for MMF to converge than GMF.  The second observation is that QCF usually outperforms MMF by a large margin, which also ascertains the reasonableness of using Quaternion for recommender systems. Compared with the simple combination in MMF, Quaternion collaborative filtering enables more complicated interactions and is less likely to cause over-fitting.

\begin{table}[t]
\small
\centering
\begin{tabular}{|l|c|c|c|c|c|c|}
\hline
\multicolumn{7}{|c|}{\textbf{Video Game}}                                          \\ \hline
            & \multicolumn{3}{c|}{Hit Ratio@k} & \multicolumn{3}{c|}{NDCG@k} \\ \hline
            & k=5      & k=10      & k=20      & k=5     & k=10    & k=20    \\ \hline
MMF         &  0.259      &     0.362     &  0.485     &  0.175      &     0.208 &  0.239    \\ \hline
$\%^{G}$ &    -0.50      &     +5.83      &  -0.39         &        -1.31 &  -0.10       &      -0.50  \\ \hline
$\%^{Q}$ &      +56.6    &    +47.0       &   +33.9        &      +63.9   &  +57.6       & +49.7        \\ \hline

\multicolumn{7}{|c|}{\textbf{Digital Music}}                                          \\ \hline
            & \multicolumn{3}{c|}{Hit Ratio@k} & \multicolumn{3}{c|}{NDCG@k} \\ \hline
            & k=5      & k=10      & k=20      & k=5     & k=10    & k=20    \\ \hline
MMF         &      0.307    &     0.406      &      0.522    &     0.206    &    0.238     & 0.267       \\ \hline
$\%^{G}$ &     -4.53     &     -2.86      &    -2.80       &    -4.22     & -3.40        &     -3.18    \\ \hline
$\%^{Q}$ &     +24.5     &  +22.3         &     +15.9      &    +29.2     &   + 27.5   &  +23.9       \\
\hline

\end{tabular}
\caption{Performance of MMF. ``$\%^{G}$" indicates the improvement of GMF over MMF. ``$\%^{Q}$" indicates the improvement of QCF over MMF.}
\end{table}

\subsubsection{Does Quaternion Neural Networks Help?}
Here, we study the impact of Quaternion neural networks. We add one Quaternion neural layer in QCF and coin it QCF+. The results are shown in Table 4. Unfortunately, there is no sign of considerable improvement with Quaternion neural networks. The increase in terms of hit ratio and NDCG is very limited. One possible explanation is that the increased amount of parameters may cause side effects to the model performance.



\begin{table}[t]
\small
\centering
\begin{tabular}{|l|c|c|c|c|c|c|}
\hline
\multicolumn{7}{|c|}{\textbf{Video Game}}                                          \\ \hline
            & \multicolumn{3}{c|}{Hit Ratio@k} & \multicolumn{3}{c|}{NDCG@k} \\ \hline
            & k=5      & k=10      & k=20      & k=5     & k=10    & k=20    \\ \hline
QCF+         &  0.409    &  0.535     & 0.657  & 0.287  &     0.326 & 0.358   \\ \hline
$\%^{Q}$ &   +0.81   &      +0.51    &   +1.17       &   +0.03   &     -0.58    &  +0.08      \\ \hline

\multicolumn{7}{|c|}{\textbf{Digital Music}}                                          \\ \hline
            & \multicolumn{3}{c|}{Hit Ratio@k} & \multicolumn{3}{c|}{NDCG@k} \\ \hline
            & k=5      & k=10      & k=20      & k=5     & k=10    & k=20    \\ \hline
QCF+         &  0.386    &   0.502      &  0.614   &  0.273    &  0.311     &   0.339    \\ \hline
$\%^{Q}$ &     +1.03    &   +1.07     &    +1.49      &   +2.55    &  +2.50 &  +2.45     \\
\hline

\end{tabular}
\caption{Performance of Quaternion neural networks based recommender systems. Here, ``$\%^{Q}$" indicates the performance improvement of QCF+ over QCF.}
\end{table}

\subsubsection{Impact of Embedding Dimension $d$}
Figure \ref{fig:parame} (a) and (b) show the effect of hyper-parameter $d$ on the performance of QCF and NeuMF. Evidently, the embedding dimension has a big effect on the model performance. We make two observations. On the one hand, our model consistently outperforms NeuMF by varying the embedding size. On the other hand, both small or large embedding dimension can pose side effects on the results. Setting $d$ to a value between $15$ to $45$ is usually preferable. The performances of both models increase largely by moving $d$ from $5$ to $15$. From then on, the performance of NeuMF fluctuates a lot while QCF maintains increasing or stable results.

\subsubsection{Comparison on Runtime} Figure \ref{fig:parame} (c) reports the runtime of all models on datasets Video Game. The runtime includes the training time of each epoch plus the test time. As can be seen, model JRL has the highest model complexity. Our model QCF only incurs a small computational cost over the neural networks based model NeuMF. Also, the difference in runtime between CCF and GMF is insignificant. Overall, the proposed approaches are very efficient and scalable.


\section{Conclusion}


In this paper, we presented a straightforward yet effective collaborative filtering model for recommendation. We move beyond real number space and explored the effectiveness of complex and quaternion representations for recommendation tasks.  Extensive experiments on six real world datasets show that our model achieves very promising results without incurring additional cost. For future work, we will investigate more advanced hypercomplex systems such as Octonion~\cite{baez2002octonions} on the recommendation tasks.


\bibliographystyle{named}
\bibliography{ijcai19}

\begin{thebibliography}{}

\bibitem[\protect\citeauthoryear{Baez}{2002}]{baez2002octonions}
John Baez.
\newblock The octonions.
\newblock {\em Bulletin of the American Mathematical Society}, 39(2):145--205,
  2002.

\bibitem[\protect\citeauthoryear{Chen \bgroup \em et al.\egroup
  }{2017}]{chen2017personalized}
Xu~Chen, Yongfeng Zhang, Qingyao Ai, Hongteng Xu, Junchi Yan, and Zheng Qin.
\newblock Personalized key frame recommendation.
\newblock In {\em SIGIR}, pages 315--324. ACM, 2017.

\bibitem[\protect\citeauthoryear{Danihelka \bgroup \em et al.\egroup
  }{2016}]{danihelka2016associative}
Ivo Danihelka, Greg Wayne, Benigno Uria, Nal Kalchbrenner, and Alex Graves.
\newblock Associative long short-term memory.
\newblock {\em arXiv preprint arXiv:1602.03032}, 2016.

\bibitem[\protect\citeauthoryear{Dziugaite and Roy}{2015}]{dziugaite2015neural}
Gintare~Karolina Dziugaite and Daniel~M Roy.
\newblock Neural network matrix factorization.
\newblock {\em arXiv preprint arXiv:1511.06443}, 2015.

\bibitem[\protect\citeauthoryear{Gaudet and Maida}{2018}]{gaudet2018deep}
Chase~J Gaudet and Anthony~S Maida.
\newblock Deep quaternion networks.
\newblock In {\em 2018 IJCNN}, pages 1--8. IEEE, 2018.

\bibitem[\protect\citeauthoryear{Hayashi and
  Shimbo}{2017}]{hayashi2017equivalence}
Katsuhiko Hayashi and Masashi Shimbo.
\newblock On the equivalence of holographic and complex embeddings for link
  prediction.
\newblock {\em arXiv preprint arXiv:1702.05563}, 2017.

\bibitem[\protect\citeauthoryear{He \bgroup \em et al.\egroup
  }{2016}]{he2016fast}
Xiangnan He, Hanwang Zhang, Min-Yen Kan, and Tat-Seng Chua.
\newblock Fast matrix factorization for online recommendation with implicit
  feedback.
\newblock In {\em SIGIR}, 2016.

\bibitem[\protect\citeauthoryear{He \bgroup \em et al.\egroup
  }{2017}]{he2017neural}
Xiangnan He, Lizi Liao, Hanwang Zhang, Liqiang Nie, Xia Hu, and Tat-Seng Chua.
\newblock Neural collaborative filtering.
\newblock In {\em WWW}, 2017.

\bibitem[\protect\citeauthoryear{Hidasi \bgroup \em et al.\egroup
  }{2015}]{hidasi2015session}
Bal{\'a}zs Hidasi, Alexandros Karatzoglou, Linas Baltrunas, and Domonkos Tikk.
\newblock Session-based recommendations with recurrent neural networks.
\newblock {\em arXiv preprint arXiv:1511.06939}, 2015.

\bibitem[\protect\citeauthoryear{Hsieh \bgroup \em et al.\egroup
  }{2017}]{hsieh2017collaborative}
Cheng-Kang Hsieh, Longqi Yang, Yin Cui, Tsung-Yi Lin, Serge Belongie, and
  Deborah Estrin.
\newblock Collaborative metric learning.
\newblock In {\em WWW}, 2017.

\bibitem[\protect\citeauthoryear{Koren \bgroup \em et al.\egroup
  }{2009}]{koren2009matrix}
Yehuda Koren, Robert Bell, and Chris Volinsky.
\newblock Matrix factorization techniques for recommender systems.
\newblock {\em Computer}, (8):30--37, 2009.

\bibitem[\protect\citeauthoryear{Koren}{2008}]{koren2008factorization}
Yehuda Koren.
\newblock Factorization meets the neighborhood: a multifaceted collaborative
  filtering model.
\newblock In {\em KDD}, 2008.

\bibitem[\protect\citeauthoryear{Parcollet \bgroup \em et al.\egroup
  }{}]{parcollet2018quaternion2}
Titouan Parcollet, Ying Zhang, Mohamed Morchid, Chiheb Trabelsi, Georges
  Linar{\`e}s, Renato De~Mori, and Yoshua Bengio.
\newblock Quaternion convolutional neural networks for end-to-end automatic
  speech recognition.
\newblock {\em arXiv preprint arXiv:1806.07789}.

\bibitem[\protect\citeauthoryear{Parcollet \bgroup \em et al.\egroup
  }{2016}]{parcollet2016quaternionnlp}
Titouan Parcollet, Mohamed Morchid, Pierre-Michel Bousquet, Richard Dufour,
  Georges Linar{\`e}s, and Renato De~Mori.
\newblock Quaternion neural networks for spoken language understanding.
\newblock In {\em 2016 IEEE SLT Workshop}, pages 362--368. IEEE, 2016.

\bibitem[\protect\citeauthoryear{Parcollet \bgroup \em et al.\egroup
  }{2018a}]{DBLP:journals/corr/abs-1811-02656}
Titouan Parcollet, Mohamed Morchid, and Georges Linar{\`{e}}s.
\newblock Quaternion convolutional neural networks for heterogeneous image
  processing.
\newblock {\em CoRR}, abs/1811.02656, 2018.

\bibitem[\protect\citeauthoryear{Parcollet \bgroup \em et al.\egroup
  }{2018b}]{Parcollet2018SpeechRW}
Titouan Parcollet, Mirco Ravanelli, Mohamed Morchid, Georges Linar{\`e}s, and
  Renato~De Mori.
\newblock Speech recognition with quaternion neural networks.
\newblock {\em CoRR}, abs/1811.09678, 2018.

\bibitem[\protect\citeauthoryear{Parcollet \bgroup \em et al.\egroup
  }{2019}]{parcollet2018quaternion}
Titouan Parcollet, Mirco Ravanelli, Mohamed Morchid, Georges Linarès, Chiheb
  Trabelsi, Renato~De Mori, and Yoshua Bengio.
\newblock Quaternion recurrent neural networks.
\newblock In {\em ICLR}, 2019.

\bibitem[\protect\citeauthoryear{Rendle \bgroup \em et al.\egroup
  }{2009}]{rendle2009bpr}
Steffen Rendle, Christoph Freudenthaler, Zeno Gantner, and Lars Schmidt-Thieme.
\newblock Bpr: Bayesian personalized ranking from implicit feedback.
\newblock In {\em UAI}, 2009.

\bibitem[\protect\citeauthoryear{Ruder}{2016}]{ruder2016overview}
Sebastian Ruder.
\newblock An overview of gradient descent optimization algorithms.
\newblock {\em arXiv preprint arXiv:1609.04747}, 2016.

\bibitem[\protect\citeauthoryear{Sedhain \bgroup \em et al.\egroup
  }{2015}]{sedhain2015autorec}
Suvash Sedhain, Aditya~Krishna Menon, Scott Sanner, and Lexing Xie.
\newblock Autorec: Autoencoders meet collaborative filtering.
\newblock In {\em WWW}, 2015.

\bibitem[\protect\citeauthoryear{Tay \bgroup \em et al.\egroup
  }{2018a}]{tay2018latent}
Yi~Tay, Luu Anh~Tuan, and Siu~Cheung Hui.
\newblock Latent relational metric learning via memory-based attention for
  collaborative ranking.
\newblock In {\em WWW 2018}, 2018.

\bibitem[\protect\citeauthoryear{Tay \bgroup \em et al.\egroup
  }{2018b}]{tay2018hermitian}
Yi~Tay, Anh~Tuan Luu, and Siu~Cheung Hui.
\newblock Hermitian co-attention networks for text matching in asymmetrical
  domains.
\newblock 2018.

\bibitem[\protect\citeauthoryear{Tay \bgroup \em et al.\egroup
  }{2018c}]{tay2018multi}
Yi~Tay, Anh~Tuan Luu, and Siu~Cheung Hui.
\newblock Multi-pointer co-attention networks for recommendation.
\newblock In {\em Proceedings of the 24th ACM SIGKDD International Conference
  on Knowledge Discovery \& Data Mining}, pages 2309--2318. ACM, 2018.

\bibitem[\protect\citeauthoryear{Tay \bgroup \em et al.\egroup
  }{2019}]{tay2019holographic}
Yi~Tay, Shuai Zhang, Anh~Tuan Luu, Siu~Cheung Hui, Lina Yao, and Tran
  Dang~Quang Vinh.
\newblock Holographic factorization machines for recommendation.
\newblock {\em AAAI}, 2019.

\bibitem[\protect\citeauthoryear{Trabelsi \bgroup \em et al.\egroup
  }{2017}]{DBLP:journals/corr/TrabelsiBSSSMRB17}
Chiheb Trabelsi, Olexa Bilaniuk, Dmitriy Serdyuk, and et~al.
\newblock Deep complex networks.
\newblock {\em CoRR}, abs/1705.09792, 2017.

\bibitem[\protect\citeauthoryear{Trouillon \bgroup \em et al.\egroup
  }{2016}]{trouillon2016complex}
Th{\'e}o Trouillon, Johannes Welbl, Sebastian Riedel, {\'E}ric Gaussier, and
  Guillaume Bouchard.
\newblock Complex embeddings for simple link prediction.
\newblock In {\em ICML}, 2016.

\bibitem[\protect\citeauthoryear{Van~den Oord \bgroup \em et al.\egroup
  }{2013}]{van2013deep}
Aaron Van~den Oord, Sander Dieleman, and Benjamin Schrauwen.
\newblock Deep content-based music recommendation.
\newblock In {\em NIPS}, pages 2643--2651, 2013.

\bibitem[\protect\citeauthoryear{Witten and
  Shragge}{2006}]{witten2006quaternion}
Ben Witten and Jeff Shragge.
\newblock Quaternion-based signal processing.
\newblock In {\em SEG Technical Program Expanded Abstracts 2006}, pages
  2862--2866. Society of Exploration Geophysicists, 2006.

\bibitem[\protect\citeauthoryear{Wu \bgroup \em et al.\egroup
  }{2017}]{wu2017recurrent}
Chao-Yuan Wu, Amr Ahmed, Alex Beutel, Alexander~J Smola, and How Jing.
\newblock Recurrent recommender networks.
\newblock In {\em WSDM}. ACM, 2017.

\bibitem[\protect\citeauthoryear{Zhang \bgroup \em et al.\egroup
  }{2017a}]{zhang2017deep}
Shuai Zhang, Lina Yao, and Aixin Sun.
\newblock Deep learning based recommender system: A survey and new
  perspectives.
\newblock {\em arXiv preprint arXiv:1707.07435}, 2017.

\bibitem[\protect\citeauthoryear{Zhang \bgroup \em et al.\egroup
  }{2017b}]{zhang2017autosvd++}
Shuai Zhang, Lina Yao, and Xiwei Xu.
\newblock Autosvd++: An efficient hybrid collaborative filtering model via
  contractive auto-encoders.
\newblock In {\em SIGIR}, pages 957--960. ACM, 2017.

\bibitem[\protect\citeauthoryear{Zhang \bgroup \em et al.\egroup
  }{2017c}]{zhang2017joint}
Yongfeng Zhang, Qingyao Ai, Xu~Chen, and W~Bruce Croft.
\newblock Joint representation learning for top-n recommendation with
  heterogeneous information sources.
\newblock In {\em CIKM}, 2017.

\bibitem[\protect\citeauthoryear{Zhang \bgroup \em et al.\egroup
  }{2018a}]{zhang2018next}
Shuai Zhang, Yi~Tay, Lina Yao, and Aixin Sun.
\newblock Next item recommendation with self-attention.
\newblock {\em arXiv preprint arXiv:1808.06414}, 2018.

\bibitem[\protect\citeauthoryear{Zhang \bgroup \em et al.\egroup
  }{2018b}]{zhang2018neurec}
Shuai Zhang, Lina Yao, Aixin Sun, Sen Wang, Guodong Long, and Manqing Dong.
\newblock Neurec: On nonlinear transformation for personalized ranking.
\newblock {\em arXiv preprint arXiv:1805.03002}, 2018.

\bibitem[\protect\citeauthoryear{Zhang \bgroup \em et al.\egroup
  }{2019a}]{zhang2019quaternion}
Shuai Zhang, Yi~Tay, Lina Yao, and Qi~Liu.
\newblock Quaternion knowledge graph embedding.
\newblock {\em arXiv preprint arXiv:1904.10281}, 2019.

\bibitem[\protect\citeauthoryear{Zhang \bgroup \em et al.\egroup
  }{2019b}]{zhang2019deeprec}
Shuai Zhang, Yi~Tay, Lina Yao, Bin Wu, and Aixin Sun.
\newblock Deeprec: An open-source toolkit for deep learning based
  recommendation.
\newblock {\em arXiv preprint arXiv:1905.10536}, 2019.

\bibitem[\protect\citeauthoryear{Zhou \bgroup \em et al.\egroup
  }{2011}]{zhou2011functional}
Ke~Zhou, Shuang-Hong Yang, and Hongyuan Zha.
\newblock Functional matrix factorizations for cold-start recommendation.
\newblock In {\em SIGIR}, 2011.

\end{thebibliography}

\end{document}